\documentclass[aps,pra,reprint,groupedaddress,amsfonts,amssymb,amsmath]{revtex4-1}

\usepackage{graphicx}
\usepackage{dcolumn}
\usepackage{bm}
\usepackage[mathlines]{lineno}
\usepackage{siunitx}
\usepackage{booktabs}
\usepackage{color}
\usepackage{multirow}
\usepackage{lipsum}

\begin{document}

\title{Dirac electron behavior and NMR evidence for topological band inversion in ZrTe$_5$}

\author{Yefan Tian}
\affiliation{Department of Physics and Astronomy, Texas A\&M University, College Station, TX 77843, USA}
\author{Nader Ghassemi}
\affiliation{Department of Physics and Astronomy, Texas A\&M University, College Station, TX 77843, USA}
\author{Joseph H. Ross, Jr.}
\email{jhross@tamu.edu}
\affiliation{Department of Physics and Astronomy, Texas A\&M University, College Station, TX 77843, USA}

\date{\today}

\begin{abstract}
We report $^{125}$Te NMR measurements of the topological quantum material ZrTe$_5$. Spin-lattice relaxation results, well-explained by a theoretical model of Dirac electron systems, reveal that the topological characteristic of ZrTe$_5$ is $T$-dependent, changing from weak topological insulator to strong topological insulator as temperature increases. Electronic structure calculations confirm this ordering, the reverse of what has been proposed. NMR results demonstrate a gapless Dirac semimetal state occurring at a Lifshitz transition temperature, $T_c=85$ K in our crystals. We demonstrate that the changes in NMR shift at $T_c$ also provide direct evidence of band inversion when the topological phase transition occurs.
\end{abstract}

\maketitle


ZrTe$_5$ has attracted great interest as an exotic quantum material due to observations such as the chiral magnetic effect \cite{li2016chiral} and 3D quantum Hall effect \cite{tang2019three}. Initially, monolayer ZrTe$_5$ was predicted to be a 2D topological insulator (TI) \cite{weng2014transition}, with bulk ZrTe$_5$ argued to be either a weak TI (WTI) or strong TI (STI) \cite{weng2014transition}, where the latter implies a more robust protection of topological surface states from disorder, along with presence of a bulk gap. It was further predicted that a topological phase transition separates these TI states \cite{manzoni2016evidence,fan2017transition} with a temperature-driven valence and conduction band inversion associated with the topological phase transition \cite{manzoni2016evidence}. The tunable nature of these phases may be important in for example generating quantum surface states as the basis for quantum computation \cite{lee2018topological}.

Since these predictions were made, the topological nature of ZrTe$_5$ has remained controversial. Angle-resolved photoemission spectroscopy (ARPES) studies \cite{li2016chiral,shen2017spectroscopic} and the observed chiral magnetic effect \cite{li2016chiral} indicate a 3D Dirac semimetal state, also suggested by infrared \cite{chen2015optical}, magneto-optical \cite{chen2015magnetoinfrared}, and transport \cite{zheng2016transport} measurements. Based on high-resolution ARPES, however, it was concluded that ZrTe$_5$ should be a 3D WTI at low temperatures \cite{moreschini2016nature}. Scanning tunneling microscopy \cite{li2016experimental,wu2016evidence} and Shubnikov-de Haas results \cite{lv2018shubnikov} also support a WTI interpretation, while other probes of the metallic surface states argued that ZrTe$_5$ is a low-$T$ STI \cite{manzoni2016evidence,manzoni2017temperature}. Regarding the topological phase transition, a recent infrared \cite{xu2018temperature} study suggested that ZrTe$_5$ transits from WTI to STI with temperature decreasing, with the Dirac semimetal state appearing at the transition, while ARPES results \cite{zhang2017electronic} have shown the gap remaining open and the sample a WTI over the measured temperature range. 

As a powerful technique, NMR has the capability of probing both Dirac electrons and orbital symmetry changes. Here, we describe $^{125}$Te NMR measurements supported by electronic structure calculations, characterizing the 3D Dirac topological nature of ZrTe$_5$. The phase transition is shown to proceed from WTI to STI with increasing temperature associated with the bulk gap closing and reopening, while direct evidence of band inversion at the topological phase transition is established based on NMR shift measurements, demonstrating a significant capability for probing quantum materials.


\begin{figure}[t]
\includegraphics[width=0.9\columnwidth]{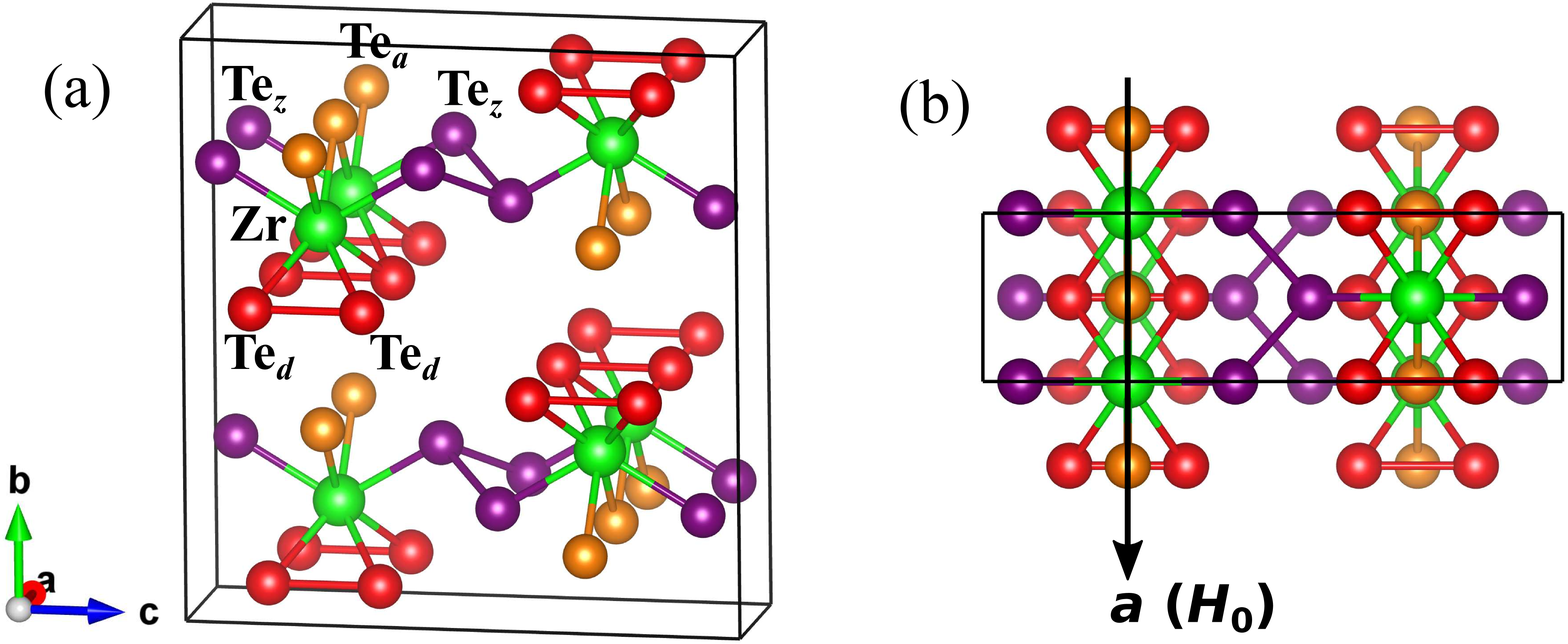}
\caption{\label{structure} (a) Crystal structure of ZrTe$_5$. Te sites include apical (Te$_a$), dimer (Te$_d$), and zigzag (Te$_z$) with occupation ratio 1:2:2. (b) $a$-$c$ plane view showing the long dimension of the needlelike crystals ($a$-axis) coinciding with the applied NMR field ($H_0$).}
\end{figure}

\begin{figure*}[t]
\includegraphics[width=1.8\columnwidth]{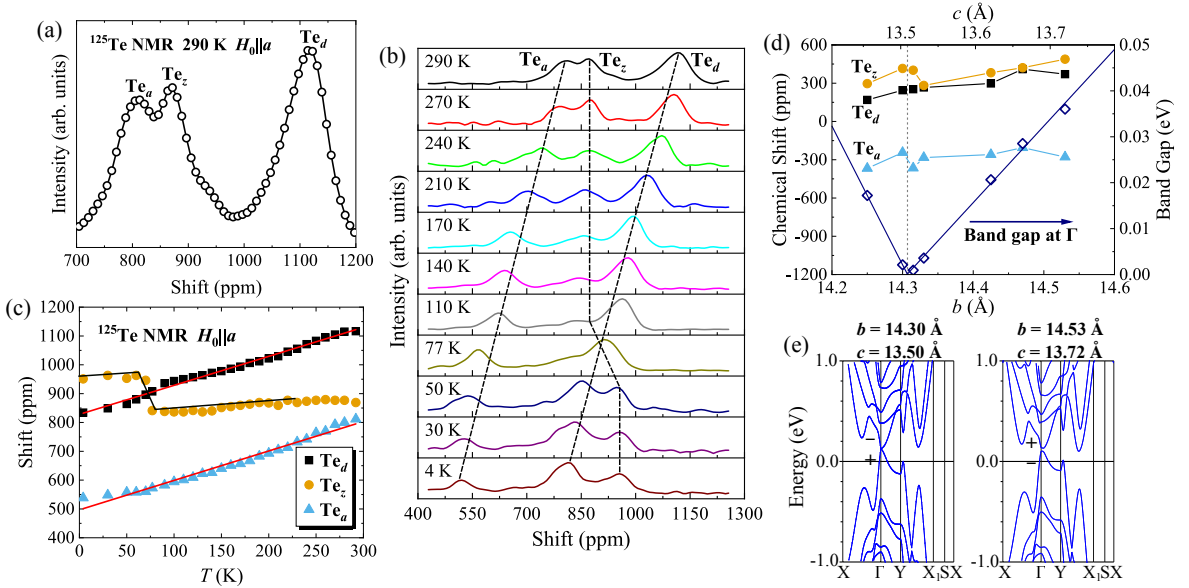}
\caption{\label{lineshapes} (a) Aligned crystal room-temperature $^{125}$Te NMR spectrum for ZrTe$_5$. (b) Temperature dependence to 4.2 K. Dashed lines: guides to the eye. (c) Fitted shift vs temperature for 3 sites. (d) Calculated band gap at $\Gamma$ and chemical shifts vs $b$ and $c$ crystal dimensions. Dashed line represents band inversion position. (e) Band structures calculated just below inversion point and for experimental lattice parameters.}
\end{figure*}

Figs.~\ref{structure}(a)-(b) show the ZrTe$_5$ crystal structure, which can be treated as ZrTe$_3$ chains connected by Te$_2$ bridging atoms. ZrTe$_5$ single crystals were prepared by chemical vapor transport (CVT). A ZrTe$_5$ precursor was prepared by reacting stoichiometric amounts of Zr (99.9\%) and Te (99.999\%) in evacuated ampules at \SI{500}{\celsius}, then mixed with 5 mg/cm$^{3}$ iodine and sealed in a quartz ampule under vacuum. The ampule was held in a 530 to \SI{470}{\celsius} gradient for one week, yielding needlelike single crystals.

Cameca SXFive microprobe measurements indicated a uniform phase ZrTe$_{5.02\pm0.02}$, equivalent within resolution to the stoichiometric composition, and larger Te content than in some other reports \cite{shahi2018bipolar,li2018giant}. No I or Hf was detected. Transport measurement showed a resistance anomaly at $\sim$125 K, typical for CVT samples. NMR experiments utilized a custom-built spectrometer at a fixed field $H_0\approx9$ T, aligning many crystals with $H_0$ parallel to $a$. Note that $H_0$$\parallel$$a$ minimizes magnetic quantum effects \cite{shahi2018bipolar}, providing a probe of an essentially unperturbed electronic structure. $^{125}$Te shifts were calibrated by aqueous Te(OH)$_6$ and adjusted for its $\delta=707$ ppm paramagnetic shift to the dimethyltelluride standard \cite{inamo1996125te}. Density functional theory (DFT) calculations were performed with WIEN2k \cite{blaha2001wien2k} using Perdew, Burke, and Ernzerhof (PBE) exchange-correlation potential, with spin-orbit coupling, a $k$-point grid of $15 \times 15 \times 4$, and atom positions from experiment \cite{fjellvaag1986structural}. Calibration of calculated $^{125}$Te chemical shifts was based on the computed ZnTe shift \cite{sirusi2016band}.



Fig.~\ref{lineshapes}(a) shows a room-temperature $^{125}$Te NMR spectrum ($I=1/2$), with peaks labeled corresponding to the three Te sites: apical (Te$_a$), dimer (Te$_d$), and zigzag (Te$_z$) [Fig.~\ref{structure}]. Fig.~\ref{lineshapes}(b) displays spectra vs temperature. Note that the number of nuclei in the expected topological edge states is negligible compared with that of the bulk so that the spectra represent the bulk. Fig.~\ref{lineshapes}(c) shows shifts obtained by fitting to three Gaussian peaks. Site assignments aided by DFT will be discussed below. 

While the Te$_d$ and Te$_a$ sites show similar behavior, steadily decreasing with temperature, Te$_z$ behaves somewhat differently, with a consistently larger line width, and about 25\% smaller spectral area than expected. With the ZrTe$_3$ chain believed to act as a rigid frame \cite{fjellvaag1986structural}, small separations and distortions of the layers apparently affect most strongly the zigzag sites causing the enhanced broadening.



\begin{figure}[t]
\includegraphics[width=0.9\columnwidth]{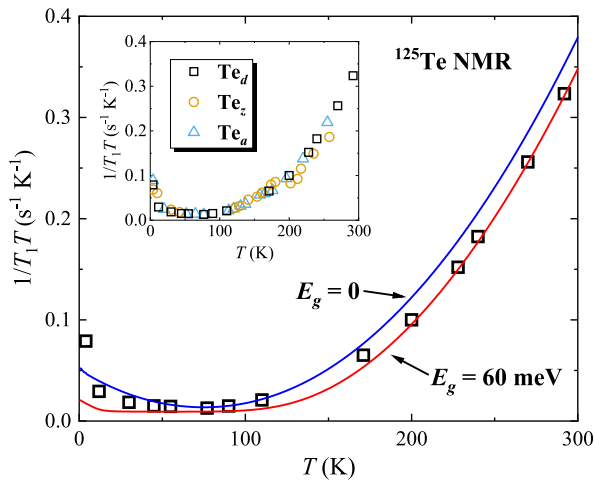}
\caption{\label{T1TvsT} $1/T_1T$ vs temperature for Te$_d$ site. Upper curve: $E_g=0$, that is, gapless Dirac semimetal in the whole temperature range, which overestimates $1/T_1T$ except near the Lifshitz $T_c$. Lower curve: $E_g=60$ meV, which matches the data far from $T_c$. Inset: $1/T_1T$ vs temperature for all sites, showing similar relaxation characteristics.}
\end{figure}

Spin-lattice relaxation, measured by inversion recovery, could be well fitted to a single exponential $M(t)=(1-Ce^{-t/T_1})M(\infty)$, giving $1/T_1T$ values shown in Fig.~\ref{T1TvsT}. The observed minimum can be regarded as indicating a density of states minimum at $E_F$ for this temperature. In metals, $1/T_1T$ is often dominated by $s$-electron Fermi contact and proportional to $g^2(E_F)$. However, with Dirac and band-edge states in ZrTe$_5$ dominated by Te $p$ states \cite{weng2014transition}, core polarization and dipolar hyperfine coupling would be expected to play more important roles. In most cases, these terms cause significant site dependence. Instead, the behavior shown in Fig.~\ref{T1TvsT} is independent of site near the minimum.

A recent model of spin-orbit-based NMR relaxation in 3D Dirac and Weyl systems accounts for this behavior very well. In this theory \cite{okvatovity2016anomalous,maebashi2018nuclear}, fluctuations in Dirac-type orbital currents are responsible for the relaxation. The orbital hyperfine interaction introduces a $1/k^2$ contribution to the momentum sum determining $1/T_1T$ \cite{okvatovity2016anomalous,okvatovity2019nuclear}, thus connecting to fluctuations that are more extended in space than the typical local contributions, explaining the site-independence. The result is a quadratic $1/T_1T$ minimum vs chemical potential ($\mu$) in the zero-$T$ limit as the Dirac point is traversed. This model was also applied to TaP \cite{okvatovity2019nuclear}, where $\mu$ pinned to a Weyl point leads to $T^2$ behavior. Here we show that this applies to the analogous case of Dirac electrons with a small gap, with $\mu$ steadily advancing through the Dirac point.

For massive Dirac fermions, the orbital contribution is \cite{maebashi2018nuclear}
\begin{multline}
\frac{1}{T_1T}=\frac{2\pi}{3}\mu_0^2\gamma_n^2e^2{c^*}^4 \\
\times \int_{-\infty}^\infty dE\bigg[-\frac{\partial f(E,\mu)}{\partial E}\bigg]\frac{g^2(E)}{E^2}\ln\frac{2(E^2-\Delta^2)}{\omega_0|E|},
\end{multline}
with $E=\pm\sqrt{{c^*}^2k^2+\Delta^2}$. In addition, $f(E,\mu)$ is the Fermi function, and $g(E)$ is the Dirac electron density of states,
\begin{equation}
g(E)=\frac{|E|\sqrt{E^2-\Delta^2}}{2\pi^2{c^*}^3}\theta(E^2-\Delta^2),
\end{equation}
with $\theta(E^2-\Delta^2)$ a step function enforcing no states in the $E_g=2\Delta$ gap. The result is
\begin{multline} \label{calc_gap}
\frac{1}{T_1T}=\frac{\mu_0^2\gamma_n^2e^2k_B\beta}{6\pi^3{c^*}^2\hbar^3} \\
\times\int\limits_{|E|\geqslant\Delta} dE\frac{(E^2-\Delta^2)\ln[2(E^2-\Delta^2)/\hbar\omega_0|E|]}{4k_BT\cosh^2[(E-\mu)/2k_BT]},
\end{multline}
where $\beta$ is an overall scale factor \cite{okvatovity2019nuclear} accounting for details of the Bloch wavefunctions.


In fitting $1/T_1T$, we assumed $\mu$ is positioned in the conduction band at low temperature, and advances through the Dirac point as $T$ increases, consistent with the observed $n$- to $p$-type change \cite{shahi2018bipolar,miller2018polycrystalline} as well as ARPES measurements \cite{zhang2017electronic}. By numerically integrating Eq.~(\ref{calc_gap}), we found that a linear decrease in $\mu$ vs $T$ gives results that agree with the higher-temperature data, but only with the gap set to $2\Delta\approx60$ meV. However, close to the minimum, the smaller curvature indicates a gap approaching zero. This is illustrated in Fig.~\ref{T1TvsT} for the Te$_d$ site with gapless and gapped ($E_g=60$ meV) cases shown by the labeled curves, with a very small $1/T_1T=0.009$ s$^{-1}$\,K$^{-1}$ term added to account for other relaxation contributions. In the calculation, we replaced ${c^*}^3$ by the product of the three experimental Fermi velocities reported by Tang \textit{et al}. \cite{tang2019three}, leading to $c^*=2.1\times10^5$ m/s. The fitting at high temperatures gives $\beta=5.6\times10^{6}$, which can be compared to $\beta=8.6\times10^{6}$ reported for the Weyl case for TaP \cite{okvatovity2019nuclear}. $\beta$ and $c^*$ appear only in the prefactor of Eq.~(\ref{calc_gap}); their variation leads to a small scaling of the overall $E_g$ results without affecting the final results in a significant way.

\begin{figure}[t]
\includegraphics[width=\columnwidth]{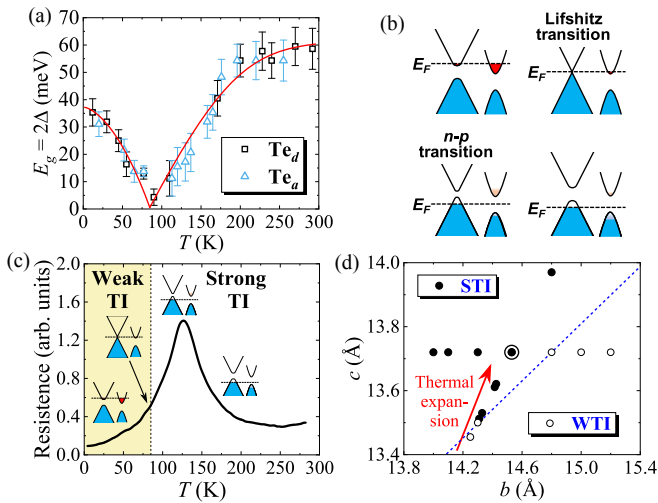}
\caption{\label{discussion} (a) Fitted band gap vs temperature obtained from $1/T_1T$ for Te$_d$ and Te$_a$ sites. Solid curves: guides to the eye. (b) Schematic of $T$-dependent chemical potential and band structure. (c) Relation between resistance and electronic structure, with WTI (shaded region) and STI as labeled. The boundary is the Dirac semimetal state. (d) Phase diagram obtained by DFT calculations. Symbols are calculated points, shaded according to band inversion. Arrow indicates experimental thermal increase of lattice parameters and boundary corresponds to the topological phase transition. Circled point: room temperature lattice parameters \cite{fjellvaag1986structural}.}
\end{figure}

Within this model, we set $\mu=\alpha(T-T_c)$ and fitted $\Delta$ vs $T$. This yielded $\alpha=-5k_B$, $T_c=85$ K, and $E_g$ vs temperature shown in Fig.~\ref{discussion}(a), clearly indicating a gap closing and opening. The closing point occurs at or very near $T_c$, where $\mu$ crosses between bands. Results for Te$_d$ and Te$_a$ sites are quite similar as shown in the plot. The Te$_z$ shift crossover prevented $T_1$ measurement in the immediate vicinity of $T_c$, although its behavior away from $T_c$ is similar to that of the other sites. These results agree well with those of Xu \textit{et al.} \cite{xu2018temperature}, although we find a larger high-$T$ gap. Also note that the fit shows that $\mu$ is positioned in the Dirac bands, rather than in the gap both above and below $T_c$.


There have been several recent reports \cite{yuan2016observation,moreschini2016nature,li2016experimental,zhang2017electronic,lv2018shubnikov,tang2019three} from Berry phase and surface imaging showing that the low-$T$ phase is a weak, rather than strong, TI. Based on these results, we can infer that the Lifshitz transition observed here corresponds to a change from WTI to STI as temperature increases. This is the reverse of what was initially proposed \cite{manzoni2016evidence,fan2017transition}, and provides a clearer picture of the topological phase transition.

DFT calculations confirm that the inversion proceeds from WTI to STI as $T$ increases. We initially scaled only $b$, and obtained DFT results equivalent to those of Ref.~\cite{manzoni2016evidence}, with a gap closing at $b=14.8$ \si{\angstrom}, and reopening with reversed parity at $\Gamma$. It was shown \cite{manzoni2016evidence} that this corresponds to a change from STI to WTI with increasing $b$. Similar results were obtained in Ref.~\cite{fan2017transition}. However, we note that the experimental thermal expansion \cite{fjellvaag1986structural} for $b$ and $c$ are essentially equal and much smaller for $a$. Thus, we examined the case of $b$ and $c$ scaled equally with $a$ held constant. The result, shown in Fig.~\ref{lineshapes}(d), is that the gap closes at $b=14.31$ \si{\angstrom}, $c=13.51$ \si{\angstrom}, for smaller instead of larger $b$. Fig.~\ref{discussion}(d) shows schematically an inferred phase boundary connecting the two inversion points identified this way. The parity of the band edges at $\Gamma$ is reversed at both inversion points. With the STI to WTI transition already demonstrated for the horizontal path in Fig.~\ref{discussion}(d), the second inversion at $\Gamma$ also requires a change of the strong $Z_2$ index and thus transition between STI and WTI \cite{fu2007topological}. 

Between 293 and 10 K, $b$ changes from 14.53 to 14.47 \si{\angstrom} \cite{fjellvaag1986structural}, with a corresponding reduction of $c$. This range does not include the predicted crossing, however use of other exchange potentials may lead to adjustment of the predicted crossing point \cite{weng2014transition}. In addition strongly $n$-type CVT crystals are reported to have smaller lattice constants \cite{shahi2018bipolar} making it appear likely that thermal expansion indeed drives the topological transition illustrated in Fig.~\ref{discussion}(d). This explains why the topological transition appears at higher temperatures in $n$-type materials with reduced lattice parameters. It also suggests that $p$-type crystals, reported to be semiconducting at all temperatures \cite{chi2017lifshitz,shahi2018bipolar}, are also STI down to zero temperature.

The difference between the Lifshitz transition temperature $T_c$ and $n$-$p$ transition temperature $T_p$ can be well explained by a two-band model \cite{chi2017lifshitz,shahi2018bipolar} as shown in Fig.~\ref{discussion}(b). While the Lifshitz transition occurs when $\mu$ passes through the Dirac point, carriers are also transferred to other minima, especially the one between $Y$ and $X$, which is nearly degenerate with the Dirac point [Fig.~\ref{lineshapes}(e)]. This is illustrated in Fig.~\ref{discussion}(c) along with the measured resistance anomaly: (i) Below $T_c$, there is $n$-type metallic behavior with $\mu$ in the Dirac and secondary conduction bands. (ii) At $T_c$, $\mu$ is at the Dirac point, which transits to a gapless semimetal state. With $\mu$ also crossing the secondary band edge, the carriers remain $n$-type due to states at the parabolic minimum. (iii) $\mu$ moves away from the secondary conduction band edge, giving the $n$-$p$ transition and the resistance anomaly. (iv) Increasing temperature produces metallic $p$-type behavior.


The carriers in the secondary minima will induce NMR Knight shifts ($K$) through their on-site spin interactions. However, based on the observed resistivity maximum, it can be estimated \cite{shahi2018bipolar} that our crystals have $n\approx1\times10^{18}$ cm$^{-3}$. For such carrier densities we estimate a contribution to $K$ which is negligible compared to the observed $T$-dependent shifts; see for example computed Te shifts for Bi$_2$Te$_3$ in Ref.~\cite{boutin2016tight}. Thus, the observed $T$-dependence must be caused by Knight shifts associated with Dirac electron spins, and/or on-site chemical shifts ($\delta$) due to the induced paramagnetic response of the valence band.

For Dirac electrons, it was recently shown \cite{okvatovity2016anomalous} that a significant dipole-generated $K$ could be expected. The limiting contribution is proportional to $\mu$ away from the Dirac point, thus linear in $T$ for the present case, with sign changing as the Dirac point is traversed. The dipole hyperfine field includes an angle-dependence which can lead to different magnitudes on each site, however, it seems likely that the nearly equivalent linear-$T$ behavior for the Te$_d$ and Te$_a$ shifts is due to the Dirac electrons, with a smaller contribution for Te$_z$. Since these contributions vanish at $T_c$ where $\mu$ goes through zero, the underlying chemical shifts can be identified from the shifts at this point.

DFT calculations of $\delta$ are shown in Fig.~\ref{lineshapes}(d), vs changes in $b$ and $c$. The shifts for Te$_d$ and Te$_z$ are nearly identical, while for Te$_a$ the result is about 500 ppm more negative. This agrees with the observed shifts at $T_c$, except for an overall negative shift. Although exchange potentials such as mBJ are expected to better reproduce the experimental shifts as opposed to PBE \cite{sirusi2016band}, the relative positions are thus rather close to what is observed. The calculated change in $\delta$ vs lattice expansion is relatively small, indicating that Dirac electrons are the dominant source for the observed linear $T$ dependence. However, the step-like change in Te$_z$ shift at the inversion point is reproduced in the calculation of $\delta$, which helps to confirm the site identification of NMR lines.

With $\delta$ associated with a local Van Vleck-type susceptibility due to partially filled Te $p$ states \cite{slichter1990principles}, the step-like change in $\delta$ also indicates a rearrangement of filled orbitals at $T_c$. The proposed band inversion was originally explained \cite{weng2014transition} in terms of a change in stabilization of $p$ orbitals on Te$_z$ and Te$_d$ sites. An associated change in orbital occupation thus will modify $\delta$, and this demonstrates that the NMR shifts in this case provide a direct measurement of the topological inversion, and thus further confirmation of the orbital interchange involved in the ZrTe$_5$ transformation. There are few techniques providing a local measurement of atomic symmetry; thus this can be a significant capability for probing quantum materials.

In summary, we explored the electronic structure and topological nature of ZrTe$_5$ using NMR techniques combined with DFT calculations. Results show that the Dirac band gap closes and reopens at a Lifshitz transition with temperature increasing, which corresponds to a topological phase transition from weak to strong topological insulator. We also show that the NMR $T_1$ results provide a very sensitive measure of the Dirac electrons involved in this transition. DFT calculations give further details about this band inversion, providing a better understanding of the topological phase transition. The observed shift change of Te$_z$ site at $T_c$ give direct evidence of the band inversion of symmetry occuring at the topological phase transition point.

\begin{acknowledgements}
The authors acknowledge helpful discussions with Bal\'{a}zs D\'{o}ra. This work was supported by the Robert A. Welch Foundation, Grant No. A-1526.
\end{acknowledgements}



\begin{thebibliography}{32}
\expandafter\ifx\csname natexlab\endcsname\relax\def\natexlab#1{#1}\fi
\expandafter\ifx\csname bibnamefont\endcsname\relax
  \def\bibnamefont#1{#1}\fi
\expandafter\ifx\csname bibfnamefont\endcsname\relax
  \def\bibfnamefont#1{#1}\fi
\expandafter\ifx\csname citenamefont\endcsname\relax
  \def\citenamefont#1{#1}\fi
\expandafter\ifx\csname url\endcsname\relax
  \def\url#1{\texttt{#1}}\fi
\expandafter\ifx\csname urlprefix\endcsname\relax\def\urlprefix{URL }\fi
\providecommand{\bibinfo}[2]{#2}
\providecommand{\eprint}[2][]{\url{#2}}

\bibitem[{\citenamefont{Li et~al.}(2016{\natexlab{a}})\citenamefont{Li,
  Kharzeev, Zhang, Huang, Pletikosi{\'c}, Fedorov, Zhong, Schneeloch, Gu, and
  Valla}}]{li2016chiral}
\bibinfo{author}{\bibfnamefont{Q.}~\bibnamefont{Li}},
  \bibinfo{author}{\bibfnamefont{D.~E.} \bibnamefont{Kharzeev}},
  \bibinfo{author}{\bibfnamefont{C.}~\bibnamefont{Zhang}},
  \bibinfo{author}{\bibfnamefont{Y.}~\bibnamefont{Huang}},
  \bibinfo{author}{\bibfnamefont{I.}~\bibnamefont{Pletikosi{\'c}}},
  \bibinfo{author}{\bibfnamefont{A.~V.} \bibnamefont{Fedorov}},
  \bibinfo{author}{\bibfnamefont{R.~D.} \bibnamefont{Zhong}},
  \bibinfo{author}{\bibfnamefont{J.~A.} \bibnamefont{Schneeloch}},
  \bibinfo{author}{\bibfnamefont{G.~D.} \bibnamefont{Gu}}, \bibnamefont{and}
  \bibinfo{author}{\bibfnamefont{T.}~\bibnamefont{Valla}},
  \bibinfo{journal}{Nat. Phys.} \textbf{\bibinfo{volume}{12}},
  \bibinfo{pages}{550} (\bibinfo{year}{2016}{\natexlab{a}}).

\bibitem[{\citenamefont{Tang et~al.}(2019)\citenamefont{Tang, Ren, Wang, Zhong,
  Schneeloch, Yang, Yang, Lee, Gu, Qiao et~al.}}]{tang2019three}
\bibinfo{author}{\bibfnamefont{F.}~\bibnamefont{Tang}},
  \bibinfo{author}{\bibfnamefont{Y.}~\bibnamefont{Ren}},
  \bibinfo{author}{\bibfnamefont{P.}~\bibnamefont{Wang}},
  \bibinfo{author}{\bibfnamefont{R.}~\bibnamefont{Zhong}},
  \bibinfo{author}{\bibfnamefont{J.}~\bibnamefont{Schneeloch}},
  \bibinfo{author}{\bibfnamefont{S.~A.} \bibnamefont{Yang}},
  \bibinfo{author}{\bibfnamefont{K.}~\bibnamefont{Yang}},
  \bibinfo{author}{\bibfnamefont{P.~A.} \bibnamefont{Lee}},
  \bibinfo{author}{\bibfnamefont{G.}~\bibnamefont{Gu}},
  \bibinfo{author}{\bibfnamefont{Z.}~\bibnamefont{Qiao}}, \bibnamefont{et~al.},
  \bibinfo{journal}{Nature} \textbf{\bibinfo{volume}{569}},
  \bibinfo{pages}{537} (\bibinfo{year}{2019}).

\bibitem[{\citenamefont{Weng et~al.}(2014)\citenamefont{Weng, Dai, and
  Fang}}]{weng2014transition}
\bibinfo{author}{\bibfnamefont{H.}~\bibnamefont{Weng}},
  \bibinfo{author}{\bibfnamefont{X.}~\bibnamefont{Dai}}, \bibnamefont{and}
  \bibinfo{author}{\bibfnamefont{Z.}~\bibnamefont{Fang}},
  \bibinfo{journal}{Phys. Rev. X} \textbf{\bibinfo{volume}{4}},
  \bibinfo{pages}{011002} (\bibinfo{year}{2014}).

\bibitem[{\citenamefont{Manzoni et~al.}(2016)\citenamefont{Manzoni,
  Gragnaniello, Aut{\`e}s, Kuhn, Sterzi, Cilento, Zacchigna, Enenkel, Vobornik,
  Barba et~al.}}]{manzoni2016evidence}
\bibinfo{author}{\bibfnamefont{G.}~\bibnamefont{Manzoni}},
  \bibinfo{author}{\bibfnamefont{L.}~\bibnamefont{Gragnaniello}},
  \bibinfo{author}{\bibfnamefont{G.}~\bibnamefont{Aut{\`e}s}},
  \bibinfo{author}{\bibfnamefont{T.}~\bibnamefont{Kuhn}},
  \bibinfo{author}{\bibfnamefont{A.}~\bibnamefont{Sterzi}},
  \bibinfo{author}{\bibfnamefont{F.}~\bibnamefont{Cilento}},
  \bibinfo{author}{\bibfnamefont{M.}~\bibnamefont{Zacchigna}},
  \bibinfo{author}{\bibfnamefont{V.}~\bibnamefont{Enenkel}},
  \bibinfo{author}{\bibfnamefont{I.}~\bibnamefont{Vobornik}},
  \bibinfo{author}{\bibfnamefont{L.}~\bibnamefont{Barba}},
  \bibnamefont{et~al.}, \bibinfo{journal}{Phys. Rev. Lett.}
  \textbf{\bibinfo{volume}{117}}, \bibinfo{pages}{237601}
  (\bibinfo{year}{2016}).

\bibitem[{\citenamefont{Fan et~al.}(2017)\citenamefont{Fan, Liang, Chen, Yao,
  and Zhou}}]{fan2017transition}
\bibinfo{author}{\bibfnamefont{Z.}~\bibnamefont{Fan}},
  \bibinfo{author}{\bibfnamefont{Q.-F.} \bibnamefont{Liang}},
  \bibinfo{author}{\bibfnamefont{Y.~B.} \bibnamefont{Chen}},
  \bibinfo{author}{\bibfnamefont{S.-H.} \bibnamefont{Yao}}, \bibnamefont{and}
  \bibinfo{author}{\bibfnamefont{J.}~\bibnamefont{Zhou}},
  \bibinfo{journal}{Sci. Rep.} \textbf{\bibinfo{volume}{7}},
  \bibinfo{pages}{45667} (\bibinfo{year}{2017}).

\bibitem[{\citenamefont{Lee et~al.}(2018)\citenamefont{Lee, Sharma,
  Lima-Sharma, Pan, and Nenoff}}]{lee2018topological}
\bibinfo{author}{\bibfnamefont{S.~R.} \bibnamefont{Lee}},
  \bibinfo{author}{\bibfnamefont{P.~A.} \bibnamefont{Sharma}},
  \bibinfo{author}{\bibfnamefont{A.~L.} \bibnamefont{Lima-Sharma}},
  \bibinfo{author}{\bibfnamefont{W.}~\bibnamefont{Pan}}, \bibnamefont{and}
  \bibinfo{author}{\bibfnamefont{T.~M.} \bibnamefont{Nenoff}},
  \bibinfo{journal}{Chem. Mater.} \textbf{\bibinfo{volume}{31}},
  \bibinfo{pages}{26} (\bibinfo{year}{2018}).

\bibitem[{\citenamefont{Shen et~al.}(2017)\citenamefont{Shen, Wang, Sun, Jiang,
  Xu, Zhang, Zhang, Lv, Yao, Chen et~al.}}]{shen2017spectroscopic}
\bibinfo{author}{\bibfnamefont{L.}~\bibnamefont{Shen}},
  \bibinfo{author}{\bibfnamefont{M.~X.} \bibnamefont{Wang}},
  \bibinfo{author}{\bibfnamefont{S.~C.} \bibnamefont{Sun}},
  \bibinfo{author}{\bibfnamefont{J.}~\bibnamefont{Jiang}},
  \bibinfo{author}{\bibfnamefont{X.}~\bibnamefont{Xu}},
  \bibinfo{author}{\bibfnamefont{T.}~\bibnamefont{Zhang}},
  \bibinfo{author}{\bibfnamefont{Q.~H.} \bibnamefont{Zhang}},
  \bibinfo{author}{\bibfnamefont{Y.~Y.} \bibnamefont{Lv}},
  \bibinfo{author}{\bibfnamefont{S.~H.} \bibnamefont{Yao}},
  \bibinfo{author}{\bibfnamefont{Y.~B.} \bibnamefont{Chen}},
  \bibnamefont{et~al.}, \bibinfo{journal}{J. Electron Spectrosc. Relat.
  Phenom.} \textbf{\bibinfo{volume}{219}}, \bibinfo{pages}{45}
  (\bibinfo{year}{2017}).

\bibitem[{\citenamefont{Chen et~al.}(2015{\natexlab{a}})\citenamefont{Chen,
  Zhang, Schneeloch, Zhang, Li, Gu, and Wang}}]{chen2015optical}
\bibinfo{author}{\bibfnamefont{R.~Y.} \bibnamefont{Chen}},
  \bibinfo{author}{\bibfnamefont{S.~J.} \bibnamefont{Zhang}},
  \bibinfo{author}{\bibfnamefont{J.~A.} \bibnamefont{Schneeloch}},
  \bibinfo{author}{\bibfnamefont{C.}~\bibnamefont{Zhang}},
  \bibinfo{author}{\bibfnamefont{Q.}~\bibnamefont{Li}},
  \bibinfo{author}{\bibfnamefont{G.~D.} \bibnamefont{Gu}}, \bibnamefont{and}
  \bibinfo{author}{\bibfnamefont{N.~L.} \bibnamefont{Wang}},
  \bibinfo{journal}{Phys. Rev. B} \textbf{\bibinfo{volume}{92}},
  \bibinfo{pages}{075107} (\bibinfo{year}{2015}{\natexlab{a}}).

\bibitem[{\citenamefont{Chen et~al.}(2015{\natexlab{b}})\citenamefont{Chen,
  Chen, Song, Schneeloch, Gu, Wang, and Wang}}]{chen2015magnetoinfrared}
\bibinfo{author}{\bibfnamefont{R.~Y.} \bibnamefont{Chen}},
  \bibinfo{author}{\bibfnamefont{Z.~G.} \bibnamefont{Chen}},
  \bibinfo{author}{\bibfnamefont{X.-Y.} \bibnamefont{Song}},
  \bibinfo{author}{\bibfnamefont{J.~A.} \bibnamefont{Schneeloch}},
  \bibinfo{author}{\bibfnamefont{G.~D.} \bibnamefont{Gu}},
  \bibinfo{author}{\bibfnamefont{F.}~\bibnamefont{Wang}}, \bibnamefont{and}
  \bibinfo{author}{\bibfnamefont{N.~L.} \bibnamefont{Wang}},
  \bibinfo{journal}{Phys. Rev. Lett.} \textbf{\bibinfo{volume}{115}},
  \bibinfo{pages}{176404} (\bibinfo{year}{2015}{\natexlab{b}}).

\bibitem[{\citenamefont{Zheng et~al.}(2016)\citenamefont{Zheng, Lu, Zhu, Ning,
  Han, Zhang, Zhang, Xi, Yang, Du et~al.}}]{zheng2016transport}
\bibinfo{author}{\bibfnamefont{G.}~\bibnamefont{Zheng}},
  \bibinfo{author}{\bibfnamefont{J.}~\bibnamefont{Lu}},
  \bibinfo{author}{\bibfnamefont{X.}~\bibnamefont{Zhu}},
  \bibinfo{author}{\bibfnamefont{W.}~\bibnamefont{Ning}},
  \bibinfo{author}{\bibfnamefont{Y.}~\bibnamefont{Han}},
  \bibinfo{author}{\bibfnamefont{H.}~\bibnamefont{Zhang}},
  \bibinfo{author}{\bibfnamefont{J.}~\bibnamefont{Zhang}},
  \bibinfo{author}{\bibfnamefont{C.}~\bibnamefont{Xi}},
  \bibinfo{author}{\bibfnamefont{J.}~\bibnamefont{Yang}},
  \bibinfo{author}{\bibfnamefont{H.}~\bibnamefont{Du}}, \bibnamefont{et~al.},
  \bibinfo{journal}{Phys. Rev. B} \textbf{\bibinfo{volume}{93}},
  \bibinfo{pages}{115414} (\bibinfo{year}{2016}).

\bibitem[{\citenamefont{Moreschini et~al.}(2016)\citenamefont{Moreschini,
  Johannsen, Berger, Denlinger, Jozwiak, Rotenberg, Kim, Bostwick, and
  Grioni}}]{moreschini2016nature}
\bibinfo{author}{\bibfnamefont{L.}~\bibnamefont{Moreschini}},
  \bibinfo{author}{\bibfnamefont{J.~C.} \bibnamefont{Johannsen}},
  \bibinfo{author}{\bibfnamefont{H.}~\bibnamefont{Berger}},
  \bibinfo{author}{\bibfnamefont{J.}~\bibnamefont{Denlinger}},
  \bibinfo{author}{\bibfnamefont{C.}~\bibnamefont{Jozwiak}},
  \bibinfo{author}{\bibfnamefont{E.}~\bibnamefont{Rotenberg}},
  \bibinfo{author}{\bibfnamefont{K.~S.} \bibnamefont{Kim}},
  \bibinfo{author}{\bibfnamefont{A.}~\bibnamefont{Bostwick}}, \bibnamefont{and}
  \bibinfo{author}{\bibfnamefont{M.}~\bibnamefont{Grioni}},
  \bibinfo{journal}{Phys. Rev. B} \textbf{\bibinfo{volume}{94}},
  \bibinfo{pages}{081101(R)} (\bibinfo{year}{2016}).

\bibitem[{\citenamefont{Li et~al.}(2016{\natexlab{b}})\citenamefont{Li, Huang,
  Lv, Zhang, Yang, Zhang, Chen, Yao, Zhou, Lu et~al.}}]{li2016experimental}
\bibinfo{author}{\bibfnamefont{X.-B.} \bibnamefont{Li}},
  \bibinfo{author}{\bibfnamefont{W.-K.} \bibnamefont{Huang}},
  \bibinfo{author}{\bibfnamefont{Y.-Y.} \bibnamefont{Lv}},
  \bibinfo{author}{\bibfnamefont{K.-W.} \bibnamefont{Zhang}},
  \bibinfo{author}{\bibfnamefont{C.-L.} \bibnamefont{Yang}},
  \bibinfo{author}{\bibfnamefont{B.-B.} \bibnamefont{Zhang}},
  \bibinfo{author}{\bibfnamefont{Y.~B.} \bibnamefont{Chen}},
  \bibinfo{author}{\bibfnamefont{S.-H.} \bibnamefont{Yao}},
  \bibinfo{author}{\bibfnamefont{J.}~\bibnamefont{Zhou}},
  \bibinfo{author}{\bibfnamefont{M.-H.} \bibnamefont{Lu}},
  \bibnamefont{et~al.}, \bibinfo{journal}{Phys. Rev. Lett.}
  \textbf{\bibinfo{volume}{116}}, \bibinfo{pages}{176803}
  (\bibinfo{year}{2016}{\natexlab{b}}).

\bibitem[{\citenamefont{Wu et~al.}(2016)\citenamefont{Wu, Ma, Nie, Zhao, Huang,
  Yin, Fu, Richard, Chen, Fang et~al.}}]{wu2016evidence}
\bibinfo{author}{\bibfnamefont{R.}~\bibnamefont{Wu}},
  \bibinfo{author}{\bibfnamefont{J.-Z.} \bibnamefont{Ma}},
  \bibinfo{author}{\bibfnamefont{S.-M.} \bibnamefont{Nie}},
  \bibinfo{author}{\bibfnamefont{L.-X.} \bibnamefont{Zhao}},
  \bibinfo{author}{\bibfnamefont{X.}~\bibnamefont{Huang}},
  \bibinfo{author}{\bibfnamefont{J.-X.} \bibnamefont{Yin}},
  \bibinfo{author}{\bibfnamefont{B.-B.} \bibnamefont{Fu}},
  \bibinfo{author}{\bibfnamefont{P.}~\bibnamefont{Richard}},
  \bibinfo{author}{\bibfnamefont{G.-F.} \bibnamefont{Chen}},
  \bibinfo{author}{\bibfnamefont{Z.}~\bibnamefont{Fang}}, \bibnamefont{et~al.},
  \bibinfo{journal}{Phys. Rev. X} \textbf{\bibinfo{volume}{6}},
  \bibinfo{pages}{021017} (\bibinfo{year}{2016}).

\bibitem[{\citenamefont{Lv et~al.}(2018)\citenamefont{Lv, Zhang, Li, Zhang, Li,
  Yao, Chen, Zhou, Zhang, Lu et~al.}}]{lv2018shubnikov}
\bibinfo{author}{\bibfnamefont{Y.-Y.} \bibnamefont{Lv}},
  \bibinfo{author}{\bibfnamefont{B.-B.} \bibnamefont{Zhang}},
  \bibinfo{author}{\bibfnamefont{X.}~\bibnamefont{Li}},
  \bibinfo{author}{\bibfnamefont{K.-W.} \bibnamefont{Zhang}},
  \bibinfo{author}{\bibfnamefont{X.-B.} \bibnamefont{Li}},
  \bibinfo{author}{\bibfnamefont{S.-H.} \bibnamefont{Yao}},
  \bibinfo{author}{\bibfnamefont{Y.~B.} \bibnamefont{Chen}},
  \bibinfo{author}{\bibfnamefont{J.}~\bibnamefont{Zhou}},
  \bibinfo{author}{\bibfnamefont{S.-T.} \bibnamefont{Zhang}},
  \bibinfo{author}{\bibfnamefont{M.-H.} \bibnamefont{Lu}},
  \bibnamefont{et~al.}, \bibinfo{journal}{Phys. Rev. B}
  \textbf{\bibinfo{volume}{97}}, \bibinfo{pages}{115137}
  (\bibinfo{year}{2018}).

\bibitem[{\citenamefont{Manzoni et~al.}(2017)\citenamefont{Manzoni, Crepaldi,
  Aut{\`e}s, Sterzi, Cilento, Akrap, Vobornik, Gragnaniello, Bugnon, Fonin
  et~al.}}]{manzoni2017temperature}
\bibinfo{author}{\bibfnamefont{G.}~\bibnamefont{Manzoni}},
  \bibinfo{author}{\bibfnamefont{A.}~\bibnamefont{Crepaldi}},
  \bibinfo{author}{\bibfnamefont{G.}~\bibnamefont{Aut{\`e}s}},
  \bibinfo{author}{\bibfnamefont{A.}~\bibnamefont{Sterzi}},
  \bibinfo{author}{\bibfnamefont{F.}~\bibnamefont{Cilento}},
  \bibinfo{author}{\bibfnamefont{A.}~\bibnamefont{Akrap}},
  \bibinfo{author}{\bibfnamefont{I.}~\bibnamefont{Vobornik}},
  \bibinfo{author}{\bibfnamefont{L.}~\bibnamefont{Gragnaniello}},
  \bibinfo{author}{\bibfnamefont{P.}~\bibnamefont{Bugnon}},
  \bibinfo{author}{\bibfnamefont{M.}~\bibnamefont{Fonin}},
  \bibnamefont{et~al.}, \bibinfo{journal}{J. Electron Spectrosc. Relat.
  Phenom.} \textbf{\bibinfo{volume}{219}}, \bibinfo{pages}{9}
  (\bibinfo{year}{2017}).

\bibitem[{\citenamefont{Xu et~al.}(2018)\citenamefont{Xu, Zhao, Marsik,
  Sheveleva, Lyzwa, Dai, Chen, Qiu, and Bernhard}}]{xu2018temperature}
\bibinfo{author}{\bibfnamefont{B.}~\bibnamefont{Xu}},
  \bibinfo{author}{\bibfnamefont{L.~X.} \bibnamefont{Zhao}},
  \bibinfo{author}{\bibfnamefont{P.}~\bibnamefont{Marsik}},
  \bibinfo{author}{\bibfnamefont{E.}~\bibnamefont{Sheveleva}},
  \bibinfo{author}{\bibfnamefont{F.}~\bibnamefont{Lyzwa}},
  \bibinfo{author}{\bibfnamefont{Y.~M.} \bibnamefont{Dai}},
  \bibinfo{author}{\bibfnamefont{G.~F.} \bibnamefont{Chen}},
  \bibinfo{author}{\bibfnamefont{X.~G.} \bibnamefont{Qiu}}, \bibnamefont{and}
  \bibinfo{author}{\bibfnamefont{C.}~\bibnamefont{Bernhard}},
  \bibinfo{journal}{Phys. Rev. Lett.} \textbf{\bibinfo{volume}{121}},
  \bibinfo{pages}{187401} (\bibinfo{year}{2018}).

\bibitem[{\citenamefont{Zhang et~al.}(2017)\citenamefont{Zhang, Wang, Yu, Liu,
  Liang, Huang, Nie, Sun, Zhang, Shen et~al.}}]{zhang2017electronic}
\bibinfo{author}{\bibfnamefont{Y.}~\bibnamefont{Zhang}},
  \bibinfo{author}{\bibfnamefont{C.}~\bibnamefont{Wang}},
  \bibinfo{author}{\bibfnamefont{L.}~\bibnamefont{Yu}},
  \bibinfo{author}{\bibfnamefont{G.}~\bibnamefont{Liu}},
  \bibinfo{author}{\bibfnamefont{A.}~\bibnamefont{Liang}},
  \bibinfo{author}{\bibfnamefont{J.}~\bibnamefont{Huang}},
  \bibinfo{author}{\bibfnamefont{S.}~\bibnamefont{Nie}},
  \bibinfo{author}{\bibfnamefont{X.}~\bibnamefont{Sun}},
  \bibinfo{author}{\bibfnamefont{Y.}~\bibnamefont{Zhang}},
  \bibinfo{author}{\bibfnamefont{B.}~\bibnamefont{Shen}}, \bibnamefont{et~al.},
  \bibinfo{journal}{Nat. Commun.} \textbf{\bibinfo{volume}{8}},
  \bibinfo{pages}{15512} (\bibinfo{year}{2017}).

\bibitem[{\citenamefont{Shahi et~al.}(2018)\citenamefont{Shahi, Singh, Sun,
  Zhao, Chen, Lv, Li, Yan, Mandrus, and Cheng}}]{shahi2018bipolar}
\bibinfo{author}{\bibfnamefont{P.}~\bibnamefont{Shahi}},
  \bibinfo{author}{\bibfnamefont{D.~J.} \bibnamefont{Singh}},
  \bibinfo{author}{\bibfnamefont{J.~P.} \bibnamefont{Sun}},
  \bibinfo{author}{\bibfnamefont{L.~X.} \bibnamefont{Zhao}},
  \bibinfo{author}{\bibfnamefont{G.~F.} \bibnamefont{Chen}},
  \bibinfo{author}{\bibfnamefont{Y.~Y.} \bibnamefont{Lv}},
  \bibinfo{author}{\bibfnamefont{J.}~\bibnamefont{Li}},
  \bibinfo{author}{\bibfnamefont{J.-Q.} \bibnamefont{Yan}},
  \bibinfo{author}{\bibfnamefont{D.~G.} \bibnamefont{Mandrus}},
  \bibnamefont{and} \bibinfo{author}{\bibfnamefont{J.-G.} \bibnamefont{Cheng}},
  \bibinfo{journal}{Phys. Rev. X} \textbf{\bibinfo{volume}{8}},
  \bibinfo{pages}{021055} (\bibinfo{year}{2018}).

\bibitem[{\citenamefont{Li et~al.}(2018)\citenamefont{Li, Zhang, Zhang, Wen,
  and Zhang}}]{li2018giant}
\bibinfo{author}{\bibfnamefont{P.}~\bibnamefont{Li}},
  \bibinfo{author}{\bibfnamefont{C.~H.} \bibnamefont{Zhang}},
  \bibinfo{author}{\bibfnamefont{J.~W.} \bibnamefont{Zhang}},
  \bibinfo{author}{\bibfnamefont{Y.}~\bibnamefont{Wen}}, \bibnamefont{and}
  \bibinfo{author}{\bibfnamefont{X.~X.} \bibnamefont{Zhang}},
  \bibinfo{journal}{Phys. Rev. B} \textbf{\bibinfo{volume}{98}},
  \bibinfo{pages}{121108(R)} (\bibinfo{year}{2018}).

\bibitem[{\citenamefont{Inamo}(1996)}]{inamo1996125te}
\bibinfo{author}{\bibfnamefont{M.}~\bibnamefont{Inamo}},
  \bibinfo{journal}{Chem. Lett.} \textbf{\bibinfo{volume}{25}},
  \bibinfo{pages}{17} (\bibinfo{year}{1996}).

\bibitem[{\citenamefont{Blaha et~al.}(2001)\citenamefont{Blaha, Schwarz,
  Madsen, Kvasnicka, and Luitz}}]{blaha2001wien2k}
\bibinfo{author}{\bibfnamefont{P.}~\bibnamefont{Blaha}},
  \bibinfo{author}{\bibfnamefont{K.}~\bibnamefont{Schwarz}},
  \bibinfo{author}{\bibfnamefont{G.~K.~H.} \bibnamefont{Madsen}},
  \bibinfo{author}{\bibfnamefont{D.}~\bibnamefont{Kvasnicka}},
  \bibnamefont{and} \bibinfo{author}{\bibfnamefont{J.}~\bibnamefont{Luitz}},
  \emph{\bibinfo{title}{{An Augmented Plane Wave$+$Local Orbitals Program for
  Calculating Crystal Properties}}} (\bibinfo{publisher}{Karlheinz Schwarz,
  Technische Universit\"{a}t Wien, Austria}, \bibinfo{year}{2001}).

\bibitem[{\citenamefont{Fjellv{\aa}g and
  Kjekshus}(1986)}]{fjellvaag1986structural}
\bibinfo{author}{\bibfnamefont{H.}~\bibnamefont{Fjellv{\aa}g}}
  \bibnamefont{and} \bibinfo{author}{\bibfnamefont{A.}~\bibnamefont{Kjekshus}},
  \bibinfo{journal}{Solid State Commun.} \textbf{\bibinfo{volume}{60}},
  \bibinfo{pages}{91} (\bibinfo{year}{1986}).

\bibitem[{\citenamefont{Sirusi et~al.}(2016)\citenamefont{Sirusi, Ballikaya,
  Chen, Uher, and Ross~Jr.}}]{sirusi2016band}
\bibinfo{author}{\bibfnamefont{A.~A.} \bibnamefont{Sirusi}},
  \bibinfo{author}{\bibfnamefont{S.}~\bibnamefont{Ballikaya}},
  \bibinfo{author}{\bibfnamefont{J.-H.} \bibnamefont{Chen}},
  \bibinfo{author}{\bibfnamefont{C.}~\bibnamefont{Uher}}, \bibnamefont{and}
  \bibinfo{author}{\bibfnamefont{J.~H.} \bibnamefont{Ross~Jr.}},
  \bibinfo{journal}{J. Phys. Chem. C} \textbf{\bibinfo{volume}{120}},
  \bibinfo{pages}{14549} (\bibinfo{year}{2016}).

\bibitem[{\citenamefont{Okv{\'a}tovity
  et~al.}(2016)\citenamefont{Okv{\'a}tovity, Simon, and
  D{\'o}ra}}]{okvatovity2016anomalous}
\bibinfo{author}{\bibfnamefont{Z.}~\bibnamefont{Okv{\'a}tovity}},
  \bibinfo{author}{\bibfnamefont{F.}~\bibnamefont{Simon}}, \bibnamefont{and}
  \bibinfo{author}{\bibfnamefont{B.}~\bibnamefont{D{\'o}ra}},
  \bibinfo{journal}{Phys. Rev. B} \textbf{\bibinfo{volume}{94}},
  \bibinfo{pages}{245141} (\bibinfo{year}{2016}).

\bibitem[{\citenamefont{Maebashi et~al.}(2018)\citenamefont{Maebashi, Hirosawa,
  Ogata, and Fukuyama}}]{maebashi2018nuclear}
\bibinfo{author}{\bibfnamefont{H.}~\bibnamefont{Maebashi}},
  \bibinfo{author}{\bibfnamefont{T.}~\bibnamefont{Hirosawa}},
  \bibinfo{author}{\bibfnamefont{M.}~\bibnamefont{Ogata}}, \bibnamefont{and}
  \bibinfo{author}{\bibfnamefont{H.}~\bibnamefont{Fukuyama}},
  \bibinfo{journal}{J. Phys. Chem. Solids} \textbf{\bibinfo{volume}{128}},
  \bibinfo{pages}{138} (\bibinfo{year}{2018}).

\bibitem[{\citenamefont{Okv{\'a}tovity
  et~al.}(2019)\citenamefont{Okv{\'a}tovity, Yasuoka, Baenitz, Simon, and
  D{\'o}ra}}]{okvatovity2019nuclear}
\bibinfo{author}{\bibfnamefont{Z.}~\bibnamefont{Okv{\'a}tovity}},
  \bibinfo{author}{\bibfnamefont{H.}~\bibnamefont{Yasuoka}},
  \bibinfo{author}{\bibfnamefont{M.}~\bibnamefont{Baenitz}},
  \bibinfo{author}{\bibfnamefont{F.}~\bibnamefont{Simon}}, \bibnamefont{and}
  \bibinfo{author}{\bibfnamefont{B.}~\bibnamefont{D{\'o}ra}},
  \bibinfo{journal}{Phys. Rev. B} \textbf{\bibinfo{volume}{99}},
  \bibinfo{pages}{115107} (\bibinfo{year}{2019}).

\bibitem[{\citenamefont{Miller et~al.}(2018)\citenamefont{Miller, Witting,
  Aydemir, Peng, Rettie, Gorai, Chung, Kanatzidis, Grayson, Stevanovi{\'c}
  et~al.}}]{miller2018polycrystalline}
\bibinfo{author}{\bibfnamefont{S.~A.} \bibnamefont{Miller}},
  \bibinfo{author}{\bibfnamefont{I.}~\bibnamefont{Witting}},
  \bibinfo{author}{\bibfnamefont{U.}~\bibnamefont{Aydemir}},
  \bibinfo{author}{\bibfnamefont{L.}~\bibnamefont{Peng}},
  \bibinfo{author}{\bibfnamefont{A.~J.~E.} \bibnamefont{Rettie}},
  \bibinfo{author}{\bibfnamefont{P.}~\bibnamefont{Gorai}},
  \bibinfo{author}{\bibfnamefont{D.~Y.} \bibnamefont{Chung}},
  \bibinfo{author}{\bibfnamefont{M.~G.} \bibnamefont{Kanatzidis}},
  \bibinfo{author}{\bibfnamefont{M.}~\bibnamefont{Grayson}},
  \bibinfo{author}{\bibfnamefont{V.}~\bibnamefont{Stevanovi{\'c}}},
  \bibnamefont{et~al.}, \bibinfo{journal}{Phys. Rev. Appl.}
  \textbf{\bibinfo{volume}{9}}, \bibinfo{pages}{014025} (\bibinfo{year}{2018}).

\bibitem[{\citenamefont{Yuan et~al.}(2016)\citenamefont{Yuan, Zhang, Liu,
  Narayan, Song, Shen, Sui, Xu, Yu, An et~al.}}]{yuan2016observation}
\bibinfo{author}{\bibfnamefont{X.}~\bibnamefont{Yuan}},
  \bibinfo{author}{\bibfnamefont{C.}~\bibnamefont{Zhang}},
  \bibinfo{author}{\bibfnamefont{Y.}~\bibnamefont{Liu}},
  \bibinfo{author}{\bibfnamefont{A.}~\bibnamefont{Narayan}},
  \bibinfo{author}{\bibfnamefont{C.}~\bibnamefont{Song}},
  \bibinfo{author}{\bibfnamefont{S.}~\bibnamefont{Shen}},
  \bibinfo{author}{\bibfnamefont{X.}~\bibnamefont{Sui}},
  \bibinfo{author}{\bibfnamefont{J.}~\bibnamefont{Xu}},
  \bibinfo{author}{\bibfnamefont{H.}~\bibnamefont{Yu}},
  \bibinfo{author}{\bibfnamefont{Z.}~\bibnamefont{An}}, \bibnamefont{et~al.},
  \bibinfo{journal}{NPG Asia Mater.} \textbf{\bibinfo{volume}{8}},
  \bibinfo{pages}{e325} (\bibinfo{year}{2016}).

\bibitem[{\citenamefont{Fu et~al.}(2007)\citenamefont{Fu, Kane, and
  Mele}}]{fu2007topological}
\bibinfo{author}{\bibfnamefont{L.}~\bibnamefont{Fu}},
  \bibinfo{author}{\bibfnamefont{C.~L.} \bibnamefont{Kane}}, \bibnamefont{and}
  \bibinfo{author}{\bibfnamefont{E.~J.} \bibnamefont{Mele}},
  \bibinfo{journal}{Phys. Rev. Lett.} \textbf{\bibinfo{volume}{98}},
  \bibinfo{pages}{106803} (\bibinfo{year}{2007}).

\bibitem[{\citenamefont{Chi et~al.}(2017)\citenamefont{Chi, Zhang, Gu,
  Kharzeev, Dai, and Li}}]{chi2017lifshitz}
\bibinfo{author}{\bibfnamefont{H.}~\bibnamefont{Chi}},
  \bibinfo{author}{\bibfnamefont{C.}~\bibnamefont{Zhang}},
  \bibinfo{author}{\bibfnamefont{G.}~\bibnamefont{Gu}},
  \bibinfo{author}{\bibfnamefont{D.~E.} \bibnamefont{Kharzeev}},
  \bibinfo{author}{\bibfnamefont{X.}~\bibnamefont{Dai}}, \bibnamefont{and}
  \bibinfo{author}{\bibfnamefont{Q.}~\bibnamefont{Li}}, \bibinfo{journal}{New
  J. Phys.} \textbf{\bibinfo{volume}{19}}, \bibinfo{pages}{015005}
  (\bibinfo{year}{2017}).

\bibitem[{\citenamefont{Boutin et~al.}(2016)\citenamefont{Boutin,
  Ram{\'i}rez-Ruiz, and Garate}}]{boutin2016tight}
\bibinfo{author}{\bibfnamefont{S.}~\bibnamefont{Boutin}},
  \bibinfo{author}{\bibfnamefont{J.}~\bibnamefont{Ram{\'i}rez-Ruiz}},
  \bibnamefont{and} \bibinfo{author}{\bibfnamefont{I.}~\bibnamefont{Garate}},
  \bibinfo{journal}{Phys. Rev. B} \textbf{\bibinfo{volume}{94}},
  \bibinfo{pages}{115204} (\bibinfo{year}{2016}).

\bibitem[{\citenamefont{Slichter}(1990)}]{slichter1990principles}
\bibinfo{author}{\bibfnamefont{C.~P.} \bibnamefont{Slichter}},
  \emph{\bibinfo{title}{Principles of Magnetic Resonance}}
  (\bibinfo{publisher}{Springer, New York}, \bibinfo{year}{1990}).

\end{thebibliography}
\end{document}